\documentclass{appolb}
\usepackage{epsfig}
\newcommand{\bk}{{\bf k}}
\newcommand{\bp}{{\bf p}}

\newcommand{\bq}{{\bf q}}

\begin{document}
\title{Nuclear model effects in neutrino-nucleus quasielastic scattering
}
\author{Chiara Maieron
\address{Istituto Nazionale di Fisica Nucleare, sezione di Catania,\\
Via Santa Sofia 64, 95123 Catania, Italy.
}}
\maketitle
\begin{abstract}
Nuclear model effects in neutrino-nucleus
quasielastic scattering are studied within
the distorted wave impulse approximation, using a relativistic shell model 
to describe the nucleus, and comparing
it with the relativistic Fermi gas.
Both charged-current and neutral-current processes are considered and,
for the neutral-current case, the uncertainties that nuclear effects may
introduce  in measurements of the axial strange form-factor
of the  nucleon are investigated.
\end{abstract}
%
%
\section{Introduction}
The interest in neutrino-nucleus scattering physics has rapidly increased
in the past few years, triggered by the rich experimental program 
aimed at studying neutrino properties which is currently under development. 
The proper interpretation
of future new data requires an accurate treatment of neutrino-nucleus
interactions, in order to minimize systematic uncertainties due to 
nuclear modeling. Additionally, the new neutrino facilities provide an 
opportunity to measure nucleon properties, such as the nucleon weak
form-factors. Such measurements will be obtained from neutrino-nucleus 
scattering data and, again, an accurate treatment of nuclear model effects
will be needed for a proper interpretation of the results in terms of 
single nucleons. Initially a Fermi gas description of the target 
nucleus was considered to be appropriate, but several studies have shown
that this description needs to be improved. To this purpose
an important guide is provided by electron-nucleus scattering, for which   
both theory and experiment have reached a very high degree of sophistication.

In this contribution we address the problem of nuclear model effects
in neutrino-nucleus quasielastic (QE) scattering, which has been studied
by many authors in the past few years 
\cite{noiCC,noiNC,noiNPAPLB,Juan,Cris,others}. We compare 
relativistic Fermi gas (RFG) calculations of cross sections with results 
obtained within a relativistic shell model (RSM), originally developed
for describing $(e,e'N)$ reactions, considering both charged-current (CC)
and neutral current (NC) processes. The former are relevant for the physics
of neutrino detectors, while the latter can be used for measuring
the nucleon axial strange form-factor. In this case, we also consider
the ratio of NC over CC cross sections.
%
\section{Formalism}

Let us consider the QE processes
\begin{eqnarray}
\nu_\mu + A \rightarrow \mu^- + p + (A-1)
\label{eq:CC}\\
\nu_\mu + A \rightarrow \nu_\mu + p + (A-1)\,,
\label{eq:NC}
\end{eqnarray}
where a neutrino of four momentum $K=(\epsilon,\bk)$ interacts
with a nucleus A, producing a final state in which
a lepton of momentum $K'=(\epsilon',\bk')$,
an emitted nucleon of momentum $P_N=(E_N,\bp_N)$ and the 
(unobserved) residual nucleus $(A-1)$ are present. 
Following standard procedures
the exclusive cross section for these processes can be written
as a contraction of a leptonic and a hadronic tensor
\begin{equation}
\frac{d\sigma}{d^3 k' d^3 p_N} 
\propto
{\eta_{\mu\nu}}{ W^{\mu\nu}}\,.
\label{eq:cross}
\end{equation}
%
The exclusive cross section is then integrated
over the momentum of the final lepton and/or nucleon
in order to obtain the observables of interest.
In the following, for the CC processes (\ref{eq:CC}) we integrate over 
the emitted proton  
and  consider the inclusive cross section
$(d\sigma/d T_\mu)$, 
where $T_\mu$ is the outgoing lepton kinetic energy. For the
NC processes (\ref{eq:NC}) we integrate over the undetectable
outgoing neutrino and consider
$(d\sigma/d T_N)$, $T_N$ being the kinetic energy of the emitted nucleon.
The leptonic tensor in Eq.~(\ref{eq:cross}) is given by
\begin{equation}
\eta_{\mu\nu} = K_\mu K'_\nu - g_{\mu\nu} K\cdot K' +  K'_\mu K_\nu\\
- i \epsilon_{\mu\nu\rho\sigma}K^\rho K^{\prime\sigma}\,.
\label{eq:lepten}
\end{equation}
The hadronic tensor $W^{\mu\nu}$
is given in general as a bilinear combination
of matrix elements of the full nuclear weak current, taken between the 
target nucleus ground state and a final
state written as a product of the residual nucleus (A-1) times
the outgoing nucleon scattering state $\phi_N$, and summed over all the states
of the residual system:
\begin{equation}
 W^{\mu\nu} = 
\sum_{(A-1)}
\langle A-1,\phi_N | \hat{J}^\mu(\bq) | A\rangle 
\langle A-1,\phi_N | \hat{J}^\nu(\bq) | A\rangle ^*
\delta(E_A + \omega - E_{A-1} -E_N)\,. 
\label{eq:hadten}
\end{equation}
We calculate  $W^{\mu\nu}$ within the impulse approximation, 
assuming (i) that the incident neutrino 
interacts with only one nucleon which is then emitted, 
while the remaining (A-1) nucleons in the target are spectators, 
(ii) that the  
nuclear current is the sum of single nucleon currents, and (iii) that
the  target and residual nuclei can be adequately described
within an independent particle model.
Under these assumptions the matrix elements contributing
to  $W^{\mu\nu}$ are greatly simplified and reduce to single nucleon
matrix elements
\begin{equation}
\langle A-1,\phi_N | \hat{J}^\mu | A\rangle 
\rightarrow\langle \phi_N |\hat{J}^\mu_{S.N.} | \psi_B \rangle \,.
\label{eq:ia}
\end{equation}
Here $\hat{J}^\mu_{S.N.}$ is the single nucleon current operator,
which we parametrize in a standard way in term of vector
and axial weak form-factors (see for 
example ~\cite{noiPREP}) and $\psi_B$ and $\phi_N$ are the
wave functions describing the initial bound nucleon and the outgoing nucleon,
respectively.
In the results presented in next section, these wave functions are calculated
using the ``Madrid-Seville'' model \cite{eep}, originally developed
for describing exclusive electron scattering reactions
and later employed extensively also for studying neutrino scattering 
\cite{noiCC,noiNC,noiNPAPLB,Juan,Cris}. 

In this model the bound nucleon wave functions
are obtained as solutions of a Dirac equation derived  within a relativistic
mean field approximation from a Lagrangian containing $\sigma$, $\omega$
and $\rho$ mesons.
Several descriptions are possible for the outgoing nucleon wave functions.
In the simplest approach, final state interactions (FSI) effects are neglected
and $\phi_N$ is given by a simple plane wave Dirac spinor (plane wave impulse
approximation, PWIA). In a more realistic approach $\phi_N$ is obtained
as the solution of a Dirac equation containing a phenomenological 
relativistic optical potential (ROP), obtained from fits of elastic 
proton-nucleus scattering data. Such potential has a real part, describing
the rescattering of the emitted nucleon, and an imaginary part, taking into
account the possibility that it is absorbed into unobserved inelastic
channels.
This description (referred to as ROP in the following)
is appropriate for the calculation of cross sections 
$(d\sigma/d T_N)$, where the final nucleon is assumed to be detected.
On the other hand, when considering inclusive cross sections 
$(d\sigma/d T_\mu)$ a selection of the single-nucleon 
knockout channel cannot be made, and 
the contribution from the inelastic channels
should be retained \cite{noiCC,Juan}. 
Within our approach, a simple way to do so is to consider
the outgoing nucleon wave functions obtained by setting to zero
the imaginary part of the ROP (real-ROP approach), 
thus taking into account the conservation
of the incident flux. 
Another possibility is to consider for $\phi_N$ the 
solutions in the continuum of the same relativistic mean field (RMF) 
equation used to obtain the nucleon bound states. Recent studies of scaling
properties of inclusive CC neutrino-nucleus QE scattering 
seem to favor the RMF approach \cite{Juan}.

We then compare our results
with those obtained within the RFG model,
(see for example
\cite{noiNC}), including in the latter also 
a phenomenological energy shift ($\omega_{sh}$),
which is introduced in studies of inclusive electron scattering in order to
get the correct position of the QE peak.
In light of current/possible experiments, we focus here
on incident energy $E_\nu=1$ GeV and on oxygen and carbon targets.
%
%
\section{Results}

\begin{figure}[t] 
\includegraphics[width=1.\columnwidth]{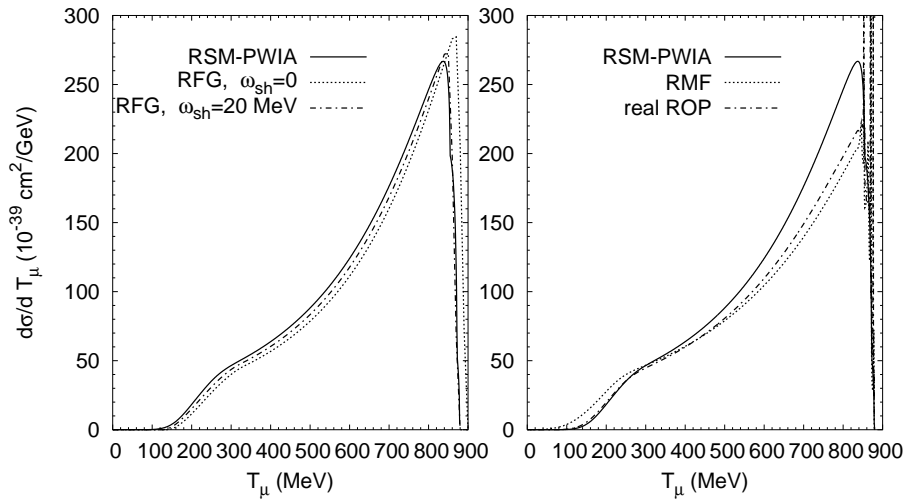}
\caption{Inclusive cross sections $d\sigma/dT_\mu$ for the
CC processes~(\ref{eq:CC}) on $^{16}O$ versus the outgoing muon 
kinetic energy $T_\mu$ for $E_\nu = 1$ GeV. The left panel
shows results in PWIA and the right panel shows the effects of FSI.
RFG cross sections are calculated for Fermi momentum $k_F=216$ MeV.}  
\label{fig:CC1000}
\end{figure} 
The differential CC cross sections 
$(d\sigma/dT_\mu)$ for the  QE scattering of muon
neutrinos on $^{16}O$ are displayed in Fig.~\ref{fig:CC1000},
as a function of the outgoing muon kinetic energy $T_\mu$ \cite{noiCC}.
In the left panel FSI effects are neglected and the RFG results
are compared with the RSM-PWIA model. 
We observe that, when the
RFG energy shift $\omega_{sh}$ is taken into account, the differences 
between RFG and RSM-PWIA are very small. The right panel
of the figure illustrates
the effects of FSI within the RSM, by comparing the
RSM-PWIA curve (solid line, the same as in the left panel) with 
the curves obtained using   
the real-ROP and
RMF approaches outlined above.  The sharp peak structure
observed at large $T_\mu$ (small energy transfer) is typical when real
potentials are used in the description of the outgoing nucleon.
The real-ROP and RMF curves, 
very close to each other, show that FSI produce a  
reduction of the cross section of about 10-15\%, which is not too
large but may have non-negligible effects in the data analysis of
experiments measuring neutrino properties.
%
\begin{figure}[t] 
\includegraphics[width=1.\columnwidth]{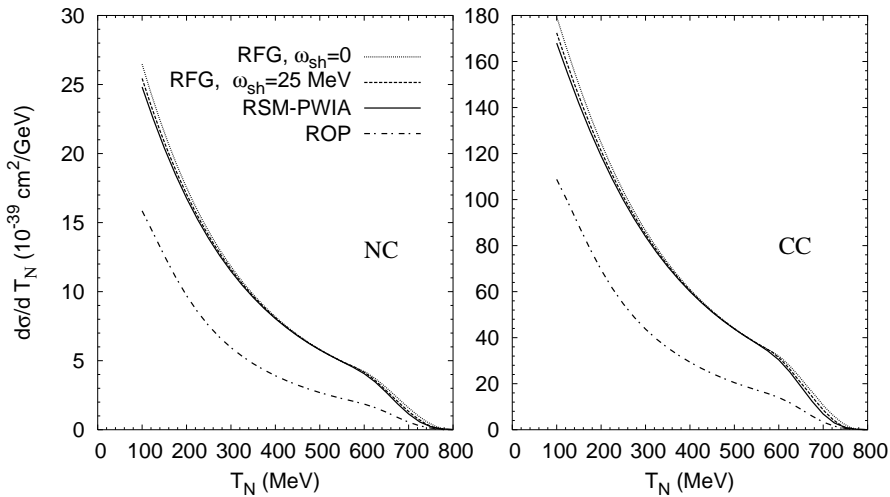}
\caption{Exclusive cross sections $d\sigma/dT_N$ for the
NC processes~(\ref{eq:NC}) (left panel) and for the CC processes 
~(\ref{eq:CC}) (right panel). The target nucleus is $^{12}C$ and the 
neutrino energy is $E_\nu= 1$ GeV. RFG cross sections are calculated for 
Fermi momentum $k_F=225$ MeV.}  
\label{fig:NC1000}
\end{figure} 

Let us now consider 
the differential cross section $(d\sigma/dT_N)$
for  $\nu_\mu$ induced proton knockout \cite{noiNC}.
Besides NC processes (left panel), we also
consider CC cross sections (right panel), which will be later used to
construct ratios of cross sections.
We see that, when FSI are neglected, the RSM and RFG
give results that almost coincide. The inclusion of FSI, which in this
case is treated using the full complex ROP, produces an important
reduction of the cross section ($\simeq 50\%$). The effects
are very similar for NC and CC processes.
%
\begin{figure}[t] 
\includegraphics[width=1.\columnwidth]{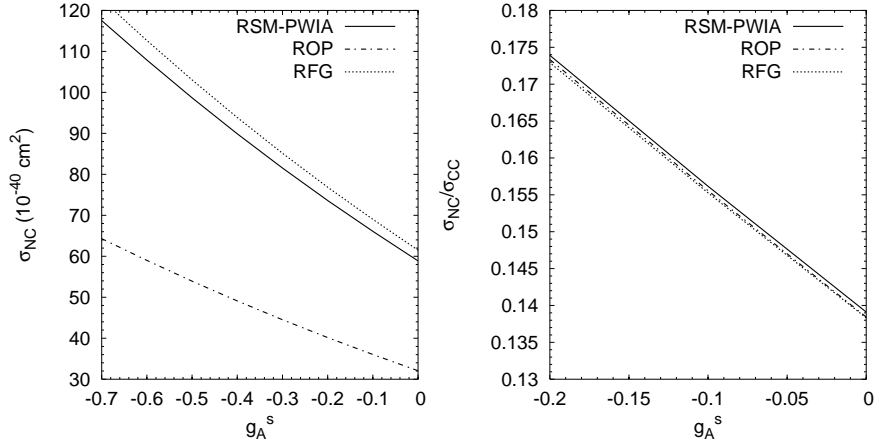}
\caption{Left panel: cross section for NC neutrino induced proton
emission integrated over the outgoing proton kinetic energy, versus
the $Q^2=0$ value of the axial strange from factor. The target nucleus
is $^{12}C$ and the incident neutrino energy is $E_\nu=1$ GeV.
Right panel: ratio of NC and CC integrated cross sections (note the
smaller range of values for $g_A^s$).} 
\label{fig:NCgas}
\end{figure} 
%

It is clear that such big effects may have a very strong impact on the
use of separate NC cross sections for measurements of the axial strange
form-factor $G_A^s$. 
This is illustrated in the left panel of Fig.~\ref{fig:NCgas}, where
the NC cross section integrated over $T_N$ is plotted as a function
of $g_A^s\equiv G_A^s(Q^2=0)$. Here a simple
dipole parametrization has been assumed for $G_A^s(Q^2)$, with the same
cut-off mass used for the non-strange axial form-factor ($M_A=1.032$ GeV).

Is is well known that, in order to extract the nucleon strange form-factors
from measurements of neutrino-nucleus cross sections, nuclear model effects
can be largely canceled by considering appropriate ratios of cross sections.
Several observables have been considered 
\cite{noiPREP,noiEL}, 
but the most realistic, from the experimental
point of view, seems to be the so called NC over CC ratio, obtained
by dividing exclusive NC cross sections by the corresponding
CC ones.
The large cancellation of nuclear model uncertainties in this ratio is 
illustrated in the  right  panel of Fig.~\ref{fig:NCgas}, where the NC/CC ratio
of integrated cross sections is plotted versus $g_A^s$.
We see that the very different curves shown in the left panel here
almost collapse on a single line, making it possible to extract the value
of  $g_A^s$, provided the experimental errors are sufficiently small.

\begin{figure}[t] 
\includegraphics[width=\columnwidth]{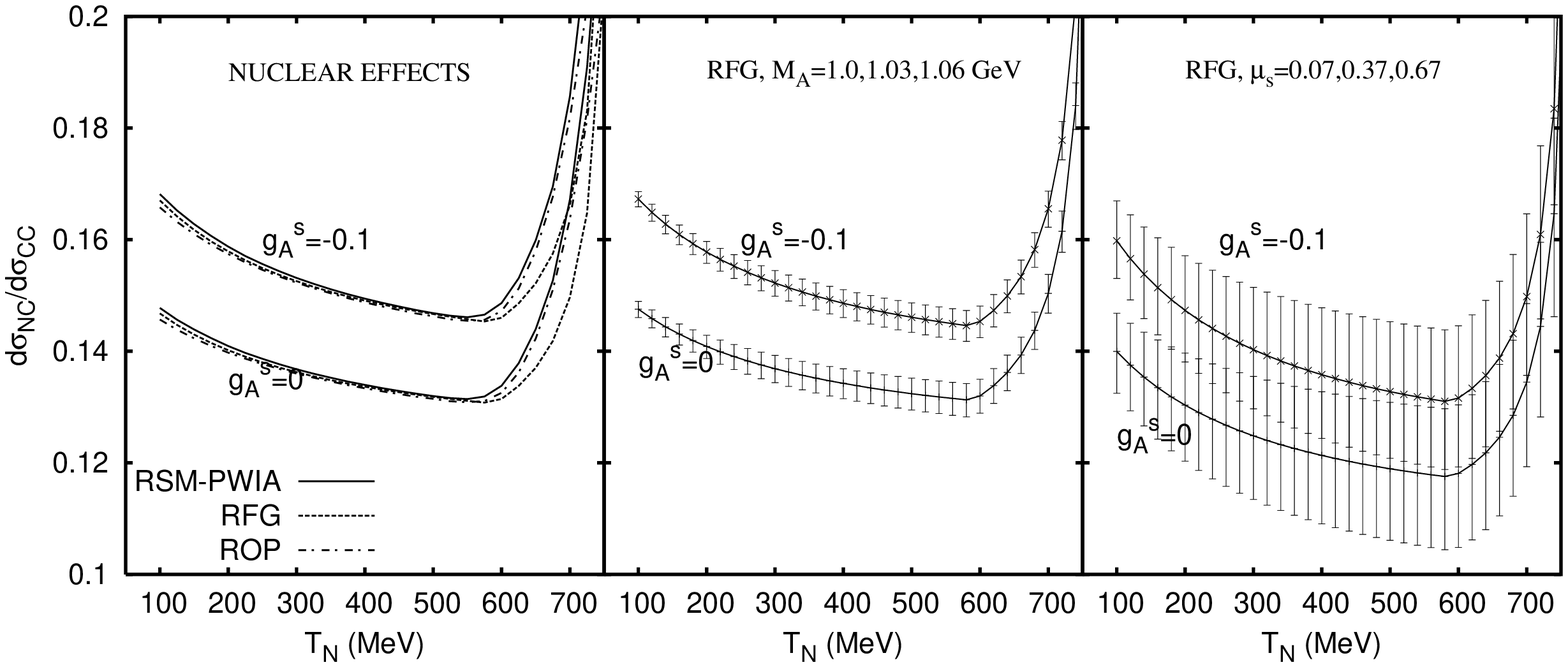}
\caption{NC over CC ratio of cross sections $d\sigma/dT_N$ for $\nu_\mu$
scattering on $^{12}C$ at $E_\nu=1$ GeV, for two values of 
$g_A^s$ as indicated. 
The left panel shows the effects of using different nuclear models.
The middle panel illustrates the uncertainties related to the value
of the axial cutoff mass. The right panel gives an example of the
possible uncertainties due to the magnetic strange form-factor $G_M^s$,
here assumed to have a dipole $Q^2$ dependence (see text).
In the middle and right panels we show only
RFG results, which are calculated for $k_F=225$ MeV
and $\omega_{sh}=0$.}  
\label{fig:NCCC}
\end{figure}

It is interesting to compare the impact of nuclear model effects on the 
determination of the axial strange form-factor $G_A^s$, 
from a measurement of the NC/CC ratio,
with other possible uncertainties, in particular those due to the 
other form-factors of the nucleon, namely the non-strange axial form-factor
$G_A$ (assumed to be measured independently in CC processes) and
the vector strange form-factors (assumed to be determined from
parity-violating electron scattering data \cite{noiPREP}).
This is done in Fig.~\ref{fig:NCCC}, where the ratio of the NC and CC 
differential cross sections  of Fig.~\ref{fig:NC1000} is plotted as a function
of the emitted nucleon energy $T_N$. 
The left panel shows the size of nuclear
model uncertainties, which are very small. 
Again, the axial form-factors $G_A$ and $G_A^s$ are parametrized assuming
a dipole dependence with the same cutoff mass $M_A$. 
The middle panel
illustrates the effects of changing $M_A$ in the range $1.00$-$1.06$ GeV.
Finally, the right panel shows the possible uncertainties due to the magnetic
strange form-factor $G_M^s$, for which we assume a dipole parametrization
$G_M^S= \mu_s/(1+Q^2/M_V^2)^2$ with $M_V^2=0.71$ $GeV^2$ 
and $\mu_s=0.37\pm 0.30$\cite{sample}, 
while we assume the electric strange form-factor $G_E^s$
to be zero. 
At present little is known about the $Q^2$ dependence
of the vector strange form-factors, and the results shown here 
are meant to be only
an illustration of the impact that the current large errors on them may have
on the NC/CC ratio. 
We see that 
the effects shown in the middle and right
panels of Fig.~\ref{fig:NCCC} 
are much larger than those due to nuclear modeling.

In conclusion,
we have calculated CC and NC neutrino-nucleus scattering cross sections
in a relativistic shell model approach, with FSI effects taken into
account within the relativistic impulse approximation.
Our results for inclusive CC QE scattering show that FSI effects, although
not being extremely large,
can still be sizable ($\simeq 10\%$) at the relatively high incident
energy $E_\nu=1$ GeV.
For exclusive NC QE processes FSI effects turn out to be very large
($\simeq 50\%$) even at high energy and may prevent a precise extraction
of the axial strange form-factor of the nucleon form separate cross sections.
However, these effects
are almost canceled when the NC/CC ratio is considered.
Within our model, nuclear model effects on ratios
turn out to be much smaller than the uncertainties introduced by the single
nucleon form-factors and can thus be considered well under control. Of 
course other uncertainties, such as those coming from non-QE contributions
to the cross sections, or due to other FSI effects not taken into account 
in the present approach, should be carefully considered.

\vspace{1.\baselineskip}
The results presented in this contribution have been obtained 
in several fruitful collaborations with W.M. Alberico, J.A. Caballero,
M.C. Mart{\'{\i}nez} and J.M. Ud{\'{\i}}as.%


\end{document}